\documentclass[pra,aps,twocolumn,showpacs]{revtex4}
\usepackage{times,epsfig,amsfonts,amsmath,amssymb,hyperref}
\newcommand{\ket}[1]{|#1\rangle}
\newcommand{\bra}[1]{\langle #1|}
\newcommand{\proj}[1]{\ket{#1}\bra{#1}}
\begin{document}

\title{Multipartite entanglement in four-qubit cluster-class states}

\author{Yan-Kui Bai and Z. D. Wang}
\email{zwang@hkucc.hku.hk}
 \affiliation{Department of Physics and
Center of Theoretical and Computational Physics, University of
Hong Kong, Pokfulam Road, Hong Kong, China}

\begin{abstract}
Based on quantitative complementarity relations (QCRs), we analyze
the multipartite correlations in four-qubit cluster-class states. It
is proven analytically that the average multipartite correlation
$E_{ms}$ is entanglement monotone. Moreover, it is also shown that
the mixed three-tangle is a correlation measure compatible with the
QCRs in this kind of quantum states. More arrestingly, with the aid
of the QCRs, a set of hierarchy entanglement measures is obtained
rigorously in the present system.
\end{abstract}


\maketitle

\section{introduction}

Entanglement, first noted by Einstein and Schr\"{o}dinger, is one of
the most important features of a many-body quantum system. Nowadays,
it is a crucial physical resource widely used in quantum information
processing (QIP), as in quantum communication \cite{eke91,ben93} and
quantum computation \cite{ben00,rau01,llb01}. Therefore, the
characterization of entanglement, especially at a quantitative
level, is fundamentally important. Compared with bipartite
entanglement, which is now well understood in many aspects, the
characterization of multipartite entanglement is still very
challenging though a lot of effort has been made (c.f.
\cite{hhh07}).

It is widely accepted that a good entanglement measure should be
non-negative, invariant under local unitary (LU) transformation, and
nonincreasing on average under local operations and classical
communications (LOCC), i.e., entanglement monotone \cite{ved97}.
Recently, based on quantitative complementarity relations (QCRs)
\cite{qcrs3}, an average multipartite correlation measure $E_{ms}$
is introduced, which was proved to satisfy the first two conditions
\cite{byw07}. From much numerical analysis, it was conjectured that
$E_{ms}$ also has the entanglement monotone property and thus may be
able to characterize the multipartite entanglement in a four-qubit
pure state \cite{byw07}. However, the analytical proof of the
conjecture is extremely difficult for a general quantum state. In
this sense, it seems  helpful to look into the conjecture in certain
cases, which, on one hand, allows us to obtain exact results, and,
on the other hand, gives us useful information beyond bipartite
entanglement.

Cluster states, which are typically multipartite entangled states,
are utilized in quantum error-correcting codes \cite{dsw02} and
tests of quantum nonlocality \cite{ogu05}. Moreover, they are also a
universal resource in one-way quantum computation \cite{rau01}. In
optical systems, a four-qubit cluster state has been prepared and
applied to the Grover search algorithm \cite{natr2,kie05} More
recently, a six-photon cluster state was also produced \cite{cyl07}.
So, in order to make better use of the cluster state, it is quite
desirable to explore quantitatively the entanglement in this kind of
system.

In this paper, we analyze the multipartite quantum correlations in
four-qubit cluster-class states. Here, by a cluster-class state , we
mean the output state of a cluster state under stochastic LOCC
(SLOCC \cite{ben01,dur00}). For this class of quantum states, we
prove exactly that the average multipartite correlation $E_{ms}$ is
entanglement monotone. Moreover, it is shown that the three- and
four-qubit correlations $t_3$ and $t_4$ are also entanglement
monotone when setting $t_3$ to be a mixed three-tangle. More
intriguingly, a set of hierarchy entanglement measures are thus
obtained rigorously in the system. The paper is organized as
follows. In Sec. II, the entanglement monotone property of
multipartite correlations in the cluster-class states is proven
exactly. In Sec. III, we address several relevant key issues and
give a brief conclusion.

\section{multipartite quantum correlations in four-qubit cluster-class states}

Before analyzing these quantum correlations, we first recall the
QCRs and the definition of average multipartite quantum correlation.
As an essential principle of quantum mechanics, complementarity
often refers to mutually exclusive properties. The quantitative
version of the complementarity relation in an $N$-qubit pure state
is also provided and formulated as \cite{qcrs3}
$\tau_{k(R_k)}+S^{2}_{k}=1$, where the linear entropy
$\tau_{k(R_k)}$ characterizes the total quantum correlation of qubit
$k$ with the remaining qubits $R_k$ and $S^{2}_{k}$ is a measure of
single-particle property. For an $N$-qubit pure state, the linear
entropy is contributed by the different levels of quantum
correlation, i.e., $\{t_2,t_{3},...,t_N\}$, in which $t_m$
represents the genuine $m$-qubit correlation for $m=2,3,...,N$
\cite{czz06,byw07}. Based on the QCRs, an average multipartite
correlation measure in a four-qubit pure state is introduced
\cite{byw07}:
\begin{equation}\label{1}
    E_{ms}(\Psi_{4})=\frac{M}{4}=\frac{M_A+M_B+M_C+M_D}{4},
\end{equation}
where $M$ is the sum of the single residual correlations and $M_k$
is defined as $M_k=\tau_{k(R_k)}-\sum_{l\in R_k} C_{kl}^2$ (here,
the square of the concurrence quantifies the two-qubit correlation).
It is conjectured that $E_{ms}$ is entanglement monotone and can
characterize the multipartite entanglement in the system. However,
the proof of this property is extreme difficult for a generic
quantum state, although a numerical analysis supports the
conjecture.

Due to the important applications in QIP, cluster states have been
paid more and more attention in recent years. As shown in Fig.1,
these states are associated with graphs where each vertex represents
a qubit prepared in the initial state $(\ket{0}+\ket{1})/\sqrt{2}$
and each edge represents a controlled phase gate applyed between two
qubits \cite{rau01}. In this paper, we will consider the
multipartite quantum correlations in four-qubit cluster-class states
that are related to the cluster states by SLOCC. In the following,
we will analyze the entanglement monotone property of the average
multipartite correlation $E_{ms}$ and the three-, and four-qubit
correlations $t_3$, and $t_4$ in this class of quantum states.

\begin{figure}
\begin{center}
\epsfig{figure=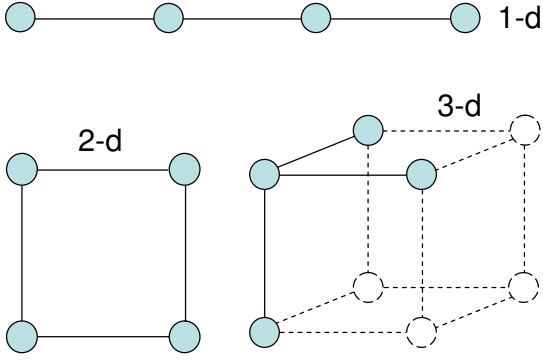,width=0.4\textwidth}
\end{center}
\caption{(Color online) The schematic graphs of four-qubit cluster
states in 1D, 2D, and 3D lattices.}
\end{figure}

\subsection{Average multipartite quantum correlation and entanglement monotone}

In one-dimensional (1D) lattices, the four-qubit cluster state can
be written as
$\ket{\mathcal{C}_{4}^{(1)}}=(\ket{0000}+\ket{0011}+\ket{1100}-\ket{1111})/2$
after LU transformation. The entanglement monotone property requires
that the correlation $E_{ms}$ does not increase on average under
LOCC. It is known that any local operation can be implemented by a
sequence of two-outcome positive operator-valued measures (POVMs)
such as $\{A_1,A_2\}$ which satisfies
$A_1^{\dagger}A_1+A_2^{\dagger}A_2=I$ \cite{dur00}. According to the
singular-value decomposition \cite{dur00}, the POVM operators can be
written as $A_1=U_1\mbox{diag}\{\alpha,\beta\}V$ and
$A_2=U_2\mbox{diag}\{\sqrt{1-\alpha^2},\sqrt{1-\beta^2}\}V$,
respectively, where $U_i$ and $V$ are unitary matrices, and $\alpha$
and $\beta$ are real numbers in range $(0,1)$. Due to the LU
invariance of the $E_{ms}$, we need only to consider the diagonal
matrices. The output state of $\ket{\mathcal{C}_4^{(1)}}$ under a
general POVM operator (i.e., the SLOCC operation) has the form
\begin{equation}\label{2}
    \ket{\Psi^{(1)}}=a\ket{0000}+b\ket{0011}+c\ket{1100}-d\ket{1111},
\end{equation}
where the normalized parameters $a,b,c$, and $d$ are complex numbers
and we refer to $\ket{\Psi^{(1)}}$ as the cluster-class state
\cite{note1}. Furthermore, since the form of this quantum state is
not changed under the next POVM, the entanglement monotone property
of $E_{ms}(\Psi^{(1)})$ will be satisfied only if the quantity is
nonincreasing under the first level of the POVM.

For the quantum state $\ket{\Psi^{(1)}}$, the two-qubit reduced
density matrix of subsystem $AB$ reads
\begin{equation}\label{3}
    \rho_{AB}=\left(%
\begin{array}{cccc}
  |a|^{2}+|b|^{2} & 0 & 0 & ac^{*}-bd^{*} \\
  0 & 0 & 0 & 0 \\
  0 & 0 & 0 & 0 \\
  a^{*}c-b^{*}d & 0 & 0 & |c|^{2}+|d|^{2} \\
\end{array}%
\right).
\end{equation}
Note that the two-qubit quantum correlation may be defined as
$t_2(\rho_{AB})=C^2(\rho_{AB})$, where the concurrence
$C(\rho_{AB})=\mbox{max}[0,(\sqrt{\lambda_1}-\sqrt{\lambda_2}-
\sqrt{\lambda_3}-\sqrt{\lambda_4})]$ with the decreasing positive
real numbers $\lambda_{i}$ being the eigenvalues of the matrix
$\rho_{AB}(\sigma_y\otimes\sigma_y)\rho_{AB}^{\ast}(\sigma_y\otimes\sigma_y)$
\cite{woo97}. After a simple calculation, we get
$C_{AB}=2|a^{*}c-b^{*}d|$. Similarly, we have
$C(\rho_{CD})=2|a^{*}b-c^{*}d|$ and $C(\rho_{ij})=0$ for other
subsystems. The linear entropy of qubit-$A$,
$\tau_{A(R_A)}(=4det\rho_{A}$) \cite{san00} can quantify the total
quantum correlation between two subsystems $A$ and $BCD$. So, the
multipartite correlation related to qubit $A$, i.e., the residual
correlation, is
\begin{equation}\label{4}
    M_A(\Psi^{(1)})=\tau_{A(R_A)}-C_{AB}^2=4|ad+bc|^2.
\end{equation}
With a similar derivation, we can obtain $M_B=M_C=M_D=M_A$, which
means that the single residual correlation $M_k(\Psi^{(1)})$ is
invariant under permutations of qubits and the average correlation
$E_{ms}(\Psi^{(1)})=M_A(\Psi^{(1)})$.

Under the POVM $\{A_1,A_2\}$ performed on the subsystem $A$, two
quantum states $\ket{\Phi_{1}^{(1)}}=A_1\ket{\Psi^{(1)}}/\sqrt{p_1}$
and $\ket{\Phi_{2}^{(1)}}=A_2\ket{\Psi^{(1)}}/\sqrt{p_2}$ are
available with probabilities
$p_i=\mbox{tr}[A_i\proj{\Psi^{(1)}}A_i^{\dagger}]$ for $i=1,2$. Note
that the linear entropy and the concurrence are invariant under
determinant one SLOCC operation (i.e., for the quantum states
$\ket{\Psi^{(1)}}$, $\ket{\Phi^{(1)}}$, and $\ket{\Phi^{(2)}}$, the
two meaures are invariant if the POVM operator satisfies
$\mbox{det}(A_i)=1$) \cite{vdd01}; we can obtain
$M_A(\Phi_1^{(1)})=\frac{\alpha^2\beta^2}{p_1^2}M_A(\Psi^{(1)})$ and
$M_A(\Phi_1^{(2)})=\frac{(1-\alpha^2)(1-\beta^2)}{p_2^2}M_A(\Psi^{(1)})$.
With a similar deduction as that in Ref. \cite{dur00}, we can derive
the following relation:
\begin{equation}\label{5}
    M_A(\Psi^{(1)})-p_1M_A(\Phi_1^{(1)})-p_2M_A(\Phi_2^{(1)})\geq 0.
\end{equation}
Combining the permutation invariance of the $M_k(\Psi^{(1)})$, we
can draw the conclusion that the single residual correlation
$M_A(\Psi^{(1)})=E_{ms}(\Psi^{(1)})$ is entanglement monotone and
can characterize the multipartite entanglement in the system.

For this kind of quantum state,  the contour plot of  $E_{ms}$
versus the non-normalized real parameters $a'$ and $d'$ is depicted
in Fig.2.1, where the parameters $b'=c'=0.5$ are fixed. In the
regions near $(a'=d'=0)$ and $(a', d'\gg0.5)$ , the multipartite
entanglement has larger values, as  the quantum state
$\ket{\Psi^{(1)}}$ tends to the Greenberger-Horne-Zeilinger (GHZ)
state. In the regions $(a'\gg b',c',d')$ and $(d'\gg a',b',c')$,
$E_{ms}$ has smaller values, as the quantum state approaches the
product state. In particular, when the real parameters $a'=d'$ and
$b'=c'$, the multipartite entanglement reaches the maximum
$E_{ms}=1$. In this case, the quantum state can be rewritten as
\begin{equation}\label{6}
    \ket{\Pi_{4}}=(\ket{00}\otimes\ket{\varphi}+\ket{11}\otimes
    \ket{\varphi^{\bot}})/\sqrt{2}
\end{equation}
where $\ket{\varphi}=(a'\ket{00}+b'\ket{11})/\sqrt{a^{'2}+b^{'2}}$
and
$\ket{\varphi^{\bot}}=b'\ket{00}-a'\ket{11}/\sqrt{a^{'2}+b^{'2}}$.
This state is the generalized Bell state, i.e., the maximal
bipartite entangled state between subsystems $AB$ and $CD$. When
$\ket{\varphi}$ is a product state, $\ket{\Pi_{4}}$ is a GHZ state.
When $\ket{\varphi}$ is a Bell state, $\ket{\Pi_{4}}$ is a cluster
state $\ket{\mathcal{C}_{4}^{(1)}}$.

\begin{figure}
\begin{center}
\epsfig{figure=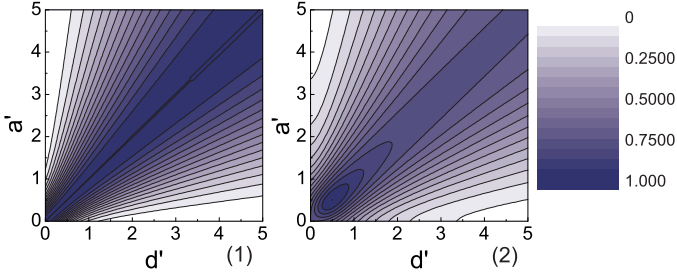,width=0.5\textwidth}
\end{center}
\caption{(Color online) Contour plots of the average multipartite
entanglement $E_{ms}$ in four-qubit cluster-class states
$\ket{\Psi^{(1)}}$ and $\ket{\Psi^{(2)}}$, where the non-normalized
parameters $a'$ and $d'$ are in the range [0,5] and the parameters
$b'=c'=0.5$ are fixed.}
\end{figure}

In two-dimensional lattices, the four-qubit cluster-class state has
the form
\begin{equation}\label{7}
    \ket{\Psi^{(2)}}=a\ket{0000}-b\ket{0111}-c\ket{1010}+d\ket{1101},
\end{equation}
where the parameters $a,b,c$, and $d$ are also complex. This kind of
quantum state is related to the box cluster state
($\ket{\mathcal{C}_{4}^{(2)}}=(\ket{0000}-\ket{0111}-\ket{1010}+\ket{1101})/2$)
via SLOCC. For the cluster-class state, we can obtain the
concurrences $C_{AC}^{2}=4(|ac|-|bd|)^2$ and $C_{ij}^{2}=0$ for the
other subsystems. Unlike in the 1D case, the single residual
correlation $M_{k}(\Psi^{(2)})$ is not permutation invariant and
does not satisfy the entanglement monotone property. As an example,
we consider the quantum state $\ket{\Psi^{(2)}}$, where the
non-normalized coefficients $a'=b'=2$, $c'=0.2$, and $d'=3$. After a
simple calculation, we have $M_A=0.5643$ and $M_C=0.2915$. Under the
POVM performed on qubit $A$ (here $\alpha=0.9$ and $\beta=0.2$), the
change of the residual correlation is $\Delta
M_C=M_C(\Psi^{(2)})-p_1M_C(\Phi_1)-p_2M_C(\Phi_2)=-0.1151$.

However, the average multipartite correlation
\begin{equation}\label{8}
    E_{ms}(\Psi^{(2)})=3(|a|^2+|c|^2)(|b|^2+|d|^2)+4|abcd|
\end{equation}
is entanglement monotone, which can be proven as follows. First, we
consider the POVM $\{A_1,A_2\}$ performed on the subsystem $A$. Due
to the LU-invariant property of the $E_{ms}$, we  need only consider
the diagonal matrices in the singular-value decomposition form, as
the output states, $\ket{\Phi_1}$ and $\ket{\Phi_2}$ are obtained
with the probabilities $p_1$ and $p_2$, respectively. The
correlation $E_{ms}(\Psi^{(2)})$ can be separated into two
components $\zeta_1=(\tau_{A(R_A)}-2C_{AC}^2)/4$ and
$\zeta_2=(\tau_{B(R_B)}+\tau_{C(R_C)}+\tau_{D(R_D)})/4$, on which
the effects are different under the POVM. The component $\zeta_1$ is
invariant under the determinant one SLOCC. With this property, we
can derive $\Delta
\zeta_1=\zeta_1(\Psi^{(2)})-p_1\zeta_1(\Phi_1)-p_2\zeta_1(\Phi_2)=
[1-\frac{\alpha^2\beta^2}{p_1}-\frac{(1-\alpha^2)(1-\beta^2)}{p_2}]\zeta_1(\Psi^{(2)})$,
where $\zeta_1(\Psi^{(2)})=(|ad|+|bc|)^2-(|ac|-|bd|)^2$ (in the
general case, this quantity is not guaranteed to be non-negative).
For the component $\zeta_2$, the change is
$\Delta\zeta_2=\zeta_2(\Psi^{(2)})-p_1\zeta_2(\Phi_1)-p_2\zeta_2(\Phi_2)=\sum_{k\neq
A}[\tau(\rho_k)-p_1\tau(\rho_k^1)-p_2\tau(\rho_k^2)]$, which is
equivalent to the changes of the linear entropies induced by the
mixed state decomposition of subsystems $\rho_{k}$ for $k=B,C,D$
\cite{note2}. After some tedious calculation, the change of the
average multipartite correlation is
\begin{eqnarray}\label{9}
    \Delta_{A} E_{ms}&=&\Delta\zeta_1+\Delta\zeta_2\nonumber\\
    &=&(\alpha^2-\beta^2)^2[4|abcd|(|a|^2+|b|^2)(|c|^2+|d|^2)\nonumber\\
    &&+3(|bc|^2-|ad|^2)^2]/p_1p_2,
\end{eqnarray}
which is obviously a non-negative number. This means that the
correlation $E_{ms}(\Psi^{(2)})$ does not increase on average under
the POVM performed on qubit $A$. But, since the quantities $\zeta_1$
and $\zeta_2$ are variant under the permutation of two qubits, we
still need to consider the POVMs performed on the subsystems $B,C$,
and $D$. After a similar analysis, we can derive the change of the
correlation under the POVM on qubit $C$ as $\Delta_{C}
E_{ms}=(\alpha^2-\beta^2)^2[4|abcd|(|a|^2+|d|^2)(|b|^2+|c|^2)
+3(|ab|^2-|cd|^2)^2]/p_1p_2$, which is also non-negative. For the
POVM on the subsystem $B$, one can separate the correlation $E_{ms}$
into two components $\kappa_1=(\tau_{B(R_B)})/4$ and
$\kappa_2=(\sum_{k\neq B}\tau_{k(R_k)}-2C_{AC}^{2})/4$ (the
non-negative property of $\kappa_2$ is guaranteed by the monogamy
relation \cite{osb06}). $\kappa_1$ is nonincreasing due to the SLOCC
invariance, and $\kappa_2$ is non-increasing because of the concave
and convex properties of the linear entropy and the concurrence,
respectively. Therefore, $E_{ms}$ is also nonincreasing under this
POVM. The case of the POVM on the subsystem $D$ is similar.
According to the above analysis, we can draw the conclusion that the
correlation $E_{ms}(\Psi^{(2)})$ is entanglement monotone and can
characterize the multipartite entanglement in the system.

In Fig.2.2, the change of $E_{ms}(\Psi^{(2)})$ with the
non-normalized real parameters $a'$ and $d'$ ($b'=c'=0.5$ are fixed)
is plotted. When $(a'\gg b',c',d')$ and $(d'\gg a',b',c')$,
$E_{ms}\approx0$ and the quantum states tend to the four-qubit
product state. When $(a',d'\approx 0)$ and $(a',d'\gg b',c')$, the
multipartite entanglement has rather large values ($E_{ms}\approx
0.75$), where the quantum state approximates to the product state of
a single-qubit state and a three-qubit GHZ state. The maximum
$E_{ms}=1$ appears at the point $a'=d'=0.5$, where the quantum state
is just the box cluster state $\ket{\mathcal{C}_4^{(2)}}$.

Finally, we address the entanglement monotone property of $E_{ms}$
in a three-dimensional cluster-class state, which is a trivial case.
This state has the form
\begin{equation}\label{10}
    \ket{\Psi^{(3)}}=a\ket{0000}+b\ket{1111},
\end{equation}
and relates to the four-qubit Greenberger-Horne-Zeilinger state via
the SLOCC operation. The quantum state $\ket{\Psi^{(3)}}$ is
invariant under the permutation of qubits and all its two-qubit
concurrences are zeros. Under the next level of the POVM, the same
properties still hold. So, the single residual correlation
$M_k=\tau_{k(R_k)}$ is entanglement monotone and satisfies
$M_A=M_B=M_C=M_D$. It is obvious that the average correlation
$E_{ms}(\Psi^{(3)})=M_k=4|ab|^2$ is also entanglement monotone and
can characterize the multipartite entanglement in the system.

\subsection{Three- and four-qubit entanglement measures}

In a four-qubit pure state $\ket{\Psi}_{ABCD}$, there are five
multipartite correlation parameters (c.f. the Venn diagram in
\cite{byw07}), i.e., one genuine four-qubit correlation
$t_4(\ket{\Psi}_{ABCD})$ and four three-qubit correlations
$t_3(\rho_{ijk})$. According to the QCRs, we have a set of equations
\cite{byw07}
\begin{equation}\label{11}
    t_4(\ket{\Psi})+\sum_{i<j\neq k}t_3(\rho_{ijk})=M_k,
\end{equation}
where $M_k$ is the single-residual correlation related to qubit $k$,
and the subscripts $i,j,k=A,B,C,D$. Note that these four equations
are unable to determine completely the five correlation parameters.
In fact, at least one additional independent relation for either
$t_3$ or $t_4$ is needed in this case.

As is known, the mixed three-tangle is a good entanglement measure
for a three-qubit mixed state; it is defined as \cite{won01}
\begin{equation}\label{12}
    \tau_{3}(\rho_{ijk})=\mbox{min}\sum_{\{p_{x},\phi_{x}\}}p_{x}\tau(\phi_{x}),
\end{equation}
where $\tau$ is the pure state three-tangle \cite{ckw00} and the
minimum runs over all the pure state decompositions of $\rho_{ijk}$.
However, it is shown in Ref. \cite{byw07} that $\tau_3$ is not
compatible with the QCRs in some specific four-qubit pure states
[for example, the quantum state
$\ket{\psi}_{ABCD}=(\ket{0000}+\ket{1011}+\ket{1101}+\ket{1110})/2$
\cite{ver02}]. So, for the cluster-class states, it is necessary to
check whether or not the $\tau_3$ can quantify correctly the $t_3$
in the QCRs. If $\tau_3$ does this, we are able to obtain the
genuine four-qubit correlation $t_4$ in terms of Eq. (11).

For the cluster-class state $\ket{\Psi^{(1)}}$ in 1D lattices, the
three-qubit reduced density matrices have the form
$\rho_{ijk}=p_1\proj{0}_i\otimes\proj{\phi}_{jk}+p_2\proj{1}_i\otimes\proj{\psi}_{jk}$,
in which $\ket{\phi}$ and $\ket{\psi}$ are two-qubit entangled
states. If one uses the mixed three-tangle to quantify the
three-qubit correlation, the relation
$t_3(\rho_{ijk})=\tau_3(\rho_{ijk})=0$ can be obtained. Substituting
this relation into Eq. (11), one can solve the genuine four-qubit
correlation $t_4=M_k=4|ad+bc|^2$. According to the analysis in Sec.
IIA, we know that the quantity $\tau_4=t_4$ satisfies all three
requirements of an entanglement measure. Therefore, for the
cluster-class state $\ket{\Psi^{(1)}}$, a set of correlation
measures $\{\tau_2,\tau_3,\tau_4\}$ which all are entanglement
monotone [we define $\tau_2(\rho_{ij})=C_{ij}^2$] can characterize
the genuine two-, three-, and four-qubit entanglement in the system.
For the cluster-state $\ket{\Psi^{(3)}}$ in 3D lattices, the case is
similar. Its three-qubit reduced density matrix is
$\rho_{ijk}=|a|^2\proj{000}+|b|^2\proj{111}$ and the corresponding
three-tangle $\tau_3$ is zero. After using $\tau_3$ to quantify the
correlation $t_3$, one can solve the correlation
$t_4=\tau_4=M_k=4|ab|^2$, which is also entanglement monotone. So,
the correlation measures $\{\tau_2,\tau_3,\tau_4\}$ can characterize
the different levels of entanglement in the cluster-class state
$\ket{\Psi^{(3)}}$.

In the cluster-class state $\ket{\Psi^{(2)}}$, the situation is
non-trivial. If one uses the mixed three-tangle to quantify the
correlation $t_3$, it is straightforward to find that
$\tau_3(\rho_{ABC})=0$ and $\tau_3(\rho_{ACD})=0$. Substituting the
two zero $t_3$s into Eq. (11), one can obtain the other three
multipartite correlations $t_4(\Psi^{(2)})=16|abcd|$,
$t_3(\rho_{ABD})=4(|ad|-|bc|)^2$, and
$t_3(\rho_{BCD})=4(|ab|-|cd|)^2$. At this stage, we need to consider
\emph{whether or not the mixed three-tangle $\tau_3$ is compatible
with the QCRs in this system and whether the correlation $t_4$ is
appropriate to characterize the genuine four-qubit entanglement}.

We first analyze the compatibility of  $\tau_3$ with the QCRs in the
system. The decomposition of $\rho_{ABD}$ into its eigenstates can
be written as
\begin{equation}\label{13}
    \rho_{ABD}=p\proj{\psi_1}+(1-p)\proj{\psi_2},
\end{equation}
where $\ket{\psi_1}=(a\ket{000}+d\ket{111})/\sqrt{p}$,
$\ket{\psi_2}=(b\ket{011}+c\ket{100})/\sqrt{1-p}$, and
$p=|a|^2+|d|^2$. It is well known that any other decomposition can
be obtained with a unitary transformation on the eigenvectors
\cite{los06}. Hence, the vectors of any decomposition of
$\rho_{ABD}$ are linear combination of $\ket{\psi_1}$ and
$\ket{\psi_2}$, i.e.,
\begin{eqnarray}\label{14}
   \ket{Z(q,\phi)}&=&\sqrt{q}\ket{\psi_1}-e^{i\phi}\sqrt{1-q}\ket{\psi_2}\\
&=&\tilde{a}\ket{000}-e^{i\phi}\tilde{b}\ket{011}-e^{i\phi}\tilde{c}\ket{100}
+\tilde{d}\ket{111},\nonumber
\end{eqnarray}
where $\tilde{a}=a\gamma$, $\tilde{b}=b\eta$, $\tilde{c}=c\eta$, and
$\tilde{d}=d\gamma$, with $\gamma=\sqrt{q/p}$ and
$\eta=\sqrt{(1-q)/(1-p)}$. For this pure state, the reduced density
matrix of qubits $AB$ is
\begin{equation}\label{15}
    \rho_{AB}(Z)=\left(%
\begin{array}{cccc}
  |\tilde{a}|^2 & 0 & -\tilde{a}\tilde{c}^{*}e^{-i\phi} & 0 \\
  0 & |\tilde{b}|^2 & 0 & -\tilde{b}\tilde{d}^{*}e^{i\phi} \\
  -\tilde{a}^{*}\tilde{c}e^{i\phi} & 0 & |\tilde{c}|^2 & 0 \\
  0 & -\tilde{b}^{*}\tilde{d}e^{-i\phi} & 0 & |\tilde{d}|^2 \\
\end{array}%
\right)
\end{equation}
and its concurrence is zero (in fact, $\rho_{AB}$ is a mix of two
product states). Similarly, for the quantum state $\rho_{AD}(Z)$, we
can obtain $C_{AD}=0$ as well. So, in any pure state decomposition
of $\rho_{ABD}$, the entanglements of subsystems $AB$ and $AD$ are
both zeros. Then, according to the definition of the mixed state
three-tangle, we have the following relation:
\begin{eqnarray}\label{16}
  \tau_3(\rho_{ABD}) &=& \mbox{min}\sum_{\{p_x,Z_x\}}p_x\tau(Z_x(q,\phi)) \nonumber\\
  &=& \mbox{min}\sum_{\{p_x,Z_x\}}p_x[\tau_{A(R_A)}^{(x)}-(C_{AB}^{(x)})^2
  -(C_{AD}^{(x)})^2]\nonumber\\
  &=&\mbox{min}\sum_{\{p_x,Z_x\}}p_x\tau_{A(R_A)}^{(x)}\nonumber\\
  &=&C_{A:BD}^2(\rho_{ABD})\nonumber\\&=&4(|ad|-|bc|)^2,
\end{eqnarray}
where we have replaced the basis $\{\ket{00},\ket{11}\}_{BD}$ with
$\{\ket{\tilde{0}},\ket{\tilde{1}}\}_{BD}$ for the calculation of
the last equation. This value coincides with the correlation
$t_3(\rho_{ABD})$ obtained using the QCRs. For the quantum state
$\rho_{BCD}$, we can get
$\tau_3(\rho_{BCD})=4(|ab|-|cd|)^2=t_3(\rho_{BCD})$ after a similar
derivation. Therefore, in the cluster-class state
$\ket{\Psi^{(2)}}$, the mixed three-tangle $\tau_3$ can quantify
correctly the correlation $t_3$ and is compatible with the QCRs.

With the QCRs, we solve the genuine four-qubit correlation
$t_4(\Psi^{(2)})=16|abcd|$, which is obviously non-negative. The
LU-invariant property is guaranteed by the corresponding property of
the correlations $M_k$ and $t_3$ in Eq. (11). Before using
$t_4(\Psi^{(2)})$ to characterize the genuine four-qubit
entanglement in the system, we should prove first that it is
entanglement monotone. Since the correlation $t_4$ is invariant
under the permutations of qubits, we only need consider the POVM
$\{A_1,A_2\}$ performed on the subsystem $A$ in which the diagonal
matrices are $\mbox{diag} \{\alpha,\beta\}$ and $\mbox{diag}
\{\sqrt{1-\alpha^2},\sqrt{1-\beta^2}\}$, respectively. After the
POVM, two output states are available with probabilities $p_1$ and
$p_2$, respectively, and the change of the correlation is $\Delta
t_4(\Psi^{(2)})=(1-\frac{\alpha^2\beta^2}{p_1}-\frac{(1-\alpha^2)
(1-\beta^2)}{p_2})t_4(\Psi^{(2)})$. Due to the non-negativity of the
two factors in $\Delta t_4$ \cite{dur00}, the correlation
$t_4(\Psi^{(2)})=\tau_4$ is entanglement monotone. Therefore, the
set of correlation measures $\{\tau_2,\tau_3,\tau_4\}$ is able to
characterize the entanglements of two, three, and four qubits in the
cluster-class state $\ket{\Psi^{(2)}}$, namely they can be good
entanglement measures for the corresponding multi-body systems.

In Fig.3, the variations of the two-, three-, and four-qubit
entanglements with the non-normalized parameters $a'$ and $b'$ are
plotted. The behaviors of $C_{AC}^2$ and $\tau_3(\rho_{ABD})$ are
the same and both attain the maximum $0.4999$ when $(a'=0,b'=0.7)$
and $(a'=0.7,b'=0)$. The value of $\tau_3(\rho_{BCD})$ tends to 1
when $(a'=b'\approx 0)$ and $(a'=b'\gg 0.5)$, because the quantum
state $\rho_{BCD}$ approximates the pure GHZ state in these regions.
The genuine four-qubit entanglement $\tau_4$ will be 1 when
$a'=b'=0.5$. At this point, the quantum state is just the box
cluster state $\ket{\mathcal{C}^{(2)}_{4}}$.

\begin{figure}
\begin{center}
\epsfig{figure=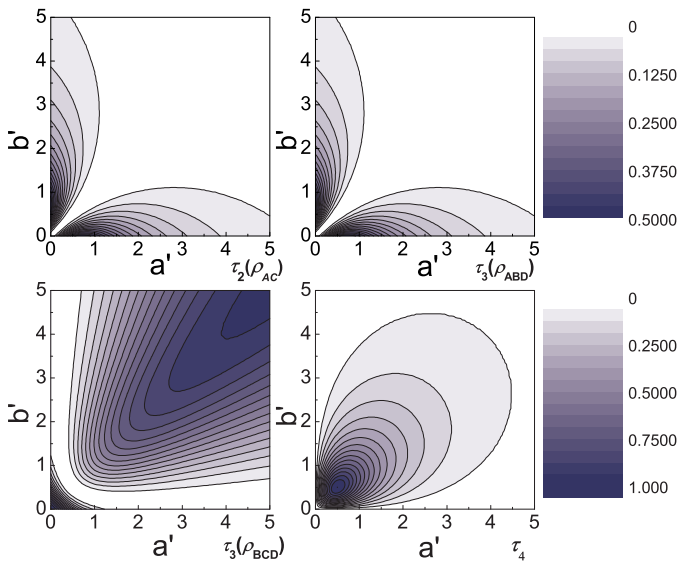,width=0.5\textwidth}
\end{center}
\caption{(Color online) Two-, three-, and four-qubit entanglement
measures versus the non-normalized real parameters $a'$ and $b'$ in the
cluster-class state
$\ket{\Psi^{(2)}}=a'\ket{0000}-b'\ket{0111}-0.5\ket{1010}+0.5\ket{1101}$.}
\end{figure}

Based on the above analysis, we conclude that not only is the mixed
three-tangle $\tau_3$ a compatible correlation measure with the QCRs
but also a set of hierarchy measures $\{\tau_2,\tau_3,\tau_4\}$ can,
respectively, quantify the two-, three-, and four-qubit entanglement
in the cluster-class states, as listed in Table I.

\begin{table}
\begin{center}
\begin{tabular}{c|c c c}
\hline\hline$\begin{array}{cc}
   & \mbox{state} \\
  \mbox{parameter} &  \\
\end{array}$ & $\ket{\Psi^{(1)}}$&  $\ket{\Psi^{(2)}}$  &  $\ket{\Psi^{(3)}}$\\\hline
$\tau_4$ & $4|ad+bc|^2$ & $16|abcd|$ & $4|ab|^2$\\
$\tau_3(\rho_{ABD})$ & $0$ & $4(|ad|-|bc|)^2$ & $0$\\
$\tau_3(\rho_{BCD})$ & $0$ &$4(|ab|-|cd|)^2$ & $0$\\
$\tau_2(\rho_{AB})$ & $4|a^{*}c-b^{*}d|^2$ & $0$ & $0$\\
$\tau_2(\rho_{AC})$ & $0$ & $4(|ac|-|bd|)^2$ & $0$\\
$\tau_2(\rho_{CD})$ & $4|a^{*}b-c^{*}d|^2$ & $0$ & $0$\\\hline\hline
\end{tabular}
\caption{Entanglement measures in different four-qubit
cluster-class states.}
\end{center}
\end{table}

\section{discussion and conclusion}

For the cluster-class state $\ket{\Psi^{(2)}}$, the single residual
correlation $M_{C}$ is not entanglement monotone as we showed in
Sec. IIA. Here, we explain the reason. This residual correlation can
be written as $M_{C}=\tau_4+\tau_3(\rho_{BCD})$ in terms of the
analysis in Sec. IIB. Although the two components are both
entanglement monotone functions under the POVMs performed on the
subsystems $B,C$, and $D$, the effects of the POVMs on the subsystem
$A$ are different from them. Due to the invariance of the qubit
permutations, $\tau_4$ is still monotone under this POVM. For the
reduced density matrix $\rho_{BCD}$, the effect of the POVM on qubit
$A$ is equivalent to a mixed state decomposition of $\rho_{BCD}$.
Because the mixed three-tangle is a convex function, the parameter
$\tau_3(\rho_{BCD})$ is nondecreasing under this POVM. Therefore,
when the decrease of $\tau_4$ is less than the increase of $\tau_3$,
the residual correlation $M_C$ will not be monotone. Just as in the
example in Sec. IIA, the changes of the three-, and four-qubit
correlations are $\Delta \tau_3(\rho_{BCD})=-0.1964$ and $\Delta
\tau_4=0.08127$, respectively, which results in $\Delta
M_C=-0.1151$. It should be pointed out that, for quantum states that
do not have three-qubit correlations under LOCC (like the
cluster-class states $\ket{\Psi^{(1)}}$ and $\ket{\Psi^{(3)}}$), the
residual correlation $M_k$ could be entanglement monotone.

In this paper, we prove analytically that $E_{ms}$ is entanglement
monotone for the four-qubit cluster-class states, and thus it can
characterize the multipartite entanglement in the system. For
general four-qubit states, $E_{ms}$ is conjectured to be
entanglement monotone according to the numerical analysis in Ref.
\cite{byw07}. Moreover, for a type of four-qubit state, numerical
analysis of Bell inequalities \cite{syu03,end05} shows a similar
property to that of $E_{ms}$, which also supports our conjecture. A
proof or disproof for an arbitrary $N$-qubit case is still awaited.
At present, we know that, in a kind of quantum state whose two-qubit
concurrences are zeros under the POVMs, the average correlation
$E_{ms}=\frac{\sum_k\tau_{k(R_k)}}{N}$ is entanglement monotone. A
trivial example is the $N$-qubit GHZ-class state
$\ket{\mathcal{G}}_N=a\ket{00\cdots0}_N+b\ket{11\cdots1}_N$. A
nontrivial example is a type of six-qubit cluster-class state
$\ket{\Psi_6}=a\ket{000000}+b\ket{000111}+c\ket{111000}-d\ket{111111}$,
where the parameters $a,b,c$, and $d$ are complex numbers; the
corresponding cluster state has been prepared recently by Lu
\emph{et al.} with a photon system \cite{cyl07}.

In the four-qubit cluster-class states, the mixed three-tangle $\tau_3$
is shown to be a compatible measure for quantifying the correlation
$t_3$ in the QCRs. With this evaluation, the genuine four-qubit
entanglement measure $\tau_4$ can be obtained. Based on this pure
cluster state entanglement, we are able to introduce a mixed state
entanglement measure by the convex roof extension \cite{uhl00},
\begin{equation}\label{17}
    \tau_4(\rho_{ABCD})=\mbox{min}\sum_{\{p_x, \phi_x^{(\mathcal{C})}\}}
    p_x\tau_4(\phi_x^{(\mathcal{C})}),
\end{equation}
where an extra restriction is that the general vector
$\ket{\phi_x^{(\mathcal{C})}}$ in the pure state decomposition has
the form of cluster-class states. As an example, we analyze the
quantum state $\rho_{ABCD}=1/2(\proj{\psi_1}+\proj{\psi_2})$, in
which $\ket{\psi_1}=(\ket{0000}+\ket{1111})/\sqrt{2}$ and
$\ket{\psi_2}=(\ket{0011}+\ket{1100})/\sqrt{2}$. The general
decomposition vector
$\ket{Z(q_k,\varphi_k)}=(\sqrt{q_k}\ket{\psi_1}-e^{i\varphi_{k}}\sqrt{1-q_k}\ket{\psi_2}$
has the form of the cluster-class state $\ket{\Psi^{(1)}}$. After
choosing $q_1=q_2=0.5$, $\varphi_{1}=0$ and $\varphi_2=\pi$,  we can
obtain $\tau_4(\rho_{ABCD})=0$ in terms of the formula in Eq. (17).
Furthermore, via the mixed state parameter $\tau_4$, one can solve
the five-qubit correlation $t_5$ with the help of the QCRs, which
can possibly be entanglement monotone in a kind of five-qubit pure
state.

In conclusion, we have explored the multipartite quantum
correlations in four-qubit cluster-class states. It is shown that
the average multipartite correlation $E_{ms}$ is entanglement
monotone in these systems,  partly supporting our previous
conjecture \cite{byw07}. Moreover, we find a set of hierarchy
measures $\{\tau_2,\tau_3,\tau_4\}$ that can characterize the
different levels of entanglement in the cluster-class states. The
entanglement monotone property of $E_{ms}$ in a general $N$-qubit
pure state is still an open problem, which is worth study in the
future.

\section*{Acknowledgments}
The authors would like to thank Dong Yang and Heng Fan for many
useful discussions and suggestions. The work was supported by the
RGC of Hong Kong under HKU Grants No. 7051/06P, 7012/06P, and 3/05C,
the URC fund of HKU, and NSF-China Grant No. 10429401.


\begin{thebibliography}{99}
\bibitem{note1}  Although the parameters $a,b,c$, and $d$ are real in the general case, we can enlarge
                 them into complex numbers without loss of generality.
\bibitem{note2}  After the POVM $\{A_1,A_2\}$, the output states of the subsystem $\rho_{B}$ will be
                 $\rho_B^i=\mbox{Tr}_{ACD}[A_i\proj{\Psi^{(2)}}A_i^\dagger]/p_i$, for
                 $i=1,2$, for which the relation $\rho_B=p_1\rho_B^1+p_2\rho_B^2$ holds. The cases for
                 $\rho_C$ and $\rho_D$ are similar.
\end{thebibliography}
\end{document}